\documentclass[pdflatex,sn-mathphys-num]{sn-jnl}

\usepackage{pdfpages}

\usepackage{graphicx}%
\usepackage{multirow}%
\usepackage{amsmath,amssymb,amsfonts}%
\usepackage{amsthm}%
\usepackage{mathrsfs}%
\usepackage[title]{appendix}%
\usepackage{xcolor}%
\usepackage{textcomp}%
\usepackage{manyfoot}%
\usepackage{booktabs}%
\usepackage{algorithm}%
\usepackage{algorithmicx}%
\usepackage{algpseudocode}%
\usepackage{listings}%
\usepackage[caption=false]{subfig}
\usepackage{siunitx}
\usepackage{braket}

\theoremstyle{thmstyleone}%
\theoremstyle{thmstyletwo}%

\theoremstyle{thmstylethree}%

\newcommand{\PrYSO}{Pr$^\textup{3+}$:Y$_2$SiO$_5$}

\begin{document}
\title{A solid-state temporally multiplexed quantum memory array at the single-photon level}
\author*[1]{\fnm{Markus} \sur{Teller}}\email{markus.teller@icfo.eu}
\equalcont{These authors contributed equally to this work.}
\author[1]{\fnm{Susana} \sur{Plascencia}}
\equalcont{These authors contributed equally to this work.}
\author[1]{\fnm{Cristina } \sur{Sastre Jachimska}}
\author[1]{\fnm{Samuele} \sur{Grandi}}
\author[1,2]{\fnm{Hugues} \sur{de Riedmatten}}
\affil[1]{\orgdiv{ICFO-Institut de Ciencies Fotoniques}, \orgname{The Barcelona Institute of Science and Technology}, \orgaddress{\city{Castelldefels (Barcelona)}, \postcode{08860}, \country{Spain}}}

\affil[2]{\orgdiv{ICREA}, \orgname{Institucio Catalana de Recerca i Estudis Avançats}, \orgaddress{\city{Barcelona}, \postcode{08015}, \country{Spain}}}

\date{\today}

\abstract{The exploitation of multimodality in different degrees of freedom is one of the most promising ways to increase the rate of heralded entanglement between distant quantum nodes. In this paper, we realize a spatially-multiplexed solid-state quantum memory array with ten individually controllable memory cells featuring on-demand read-out and temporal multiplexing. By combining spatial and temporal multiplexing, we store weak coherent pulses at the single-photon level in up to $250$ spatio-temporal modes, with an average signal-to-noise ratio of $10(2)$. We perform a thorough characterization of the whole system including its multiplexing and demultiplexing stage. We verify that the memory array exhibits low cross talk even at the single-photon level. The measured performance indicates readiness for storing non-classical states and promises a speed-up in entanglement distribution rates.}

\maketitle


\section{Introduction}

Distribution of quantum information over long distance requires quantum repeaters dispersed between quantum network nodes to counteract losses in the quantum channels~\cite{Kimble2008,Briegel1998,Sangouard2011,Wehner2018}. Quantum repeater protocols use shared entanglement between distant repeater nodes as a fundamental resource for the transfer of quantum information~\cite{Duan2001}, and thus the ability to distribute entanglement at high rates and fidelity is vital for a future quantum internet. However, a crucial challenge is imposed on the rate of heralding entanglement between distant nodes: the repetition rate is limited by the travel time of the photon through which entanglement is established and by the travel time of the classical signal heralding a successful entanglement generation~\cite{Simon2007}.  

Quantum repeaters based on multi-mode quantum memories overcome this limitation as successive entanglement attempts are generated in a multiplexed fashion, hence without waiting for the heralding signal to return~\cite{Simon2007}. Promising candidates for multiplexed quantum repeater nodes are based on atomic ensembles, either in atomic gases or in the solid-state with rare-earth doped solids. Furthermore, multiplexing has also been demonstrated recently with several trapped ions in a cavity~\cite{Krutyanskiy2024}. 
Cold atomic ensembles have been used so far mostly for spatial multiplexing with demonstrations of quantum memory arrays and wavevector multiplexing~\cite{Lan2009,Nicolas2014,Tian2017,Pu2017,Parniak2017, Li2020a, Liu2023b}. A few experiments have also shown a limited temporal multiplexing capacity~\cite{Hosseini2011, Heller2020} and conversion capabilities from time to spatial degrees~\cite{Zhang2024}, including hybrids of angular momentum and spatial modes~\cite{Zhang2024a}. 

In contrast, rare-earth doped solids allow intrinsic temporal multiplexing using the atomic frequency comb scheme (AFC)~\cite{Afzelius2009,Lago-Rivera2021,Businger2022}, as well as frequency multiplexing~\cite{Sinclair2014,Seri2019,Wei2024}, by taking advantage of the static inhomogeneous broadening of the optical transition. Another interesting aspect is that the optical depth within the crystal is constant, contrary to cold atomic ensembles, which could allow in principle uniform efficiencies across the crystal. However, so far, spatial multiplexing has not been much studied in rare-earth doped solids, although a few experiments have shown a couple of spatial modes~\cite{Gundogan2012} or a combination of all three degrees of freedom~\cite{Yang2018}. Beyond their large multiplexing capabilities~\cite{Ortu2022b}, quantum memories based on rare-earth doped solids have shown high storage and retrieval efficiencies~\cite{Hedges2010, Duranti2024}, on-demand storage of qubits with high-fidelity~\cite{Gundogan2015} and long storage times~\cite{Ortu2022a}. In addition, recent experiments with rare-earth-doped crystals demonstrated light-matter entanglement~\cite{Ferguson2016, Kutluer2019, Rakonjac2021}, matter-matter entanglement~\cite{,Lago-Rivera2021,Liu2021} and quantum teleportation~\cite{Bussieres2014,Lago-Rivera2023}, thus implementing fundamental building blocks for a quantum network.

Most experiments demonstrating large multiplexing in RE-doped crystals used storage in excited states with fixed storage times~\cite{Businger2022, Wei2024}. However, several protocols require on-demand read-out that allows retrieval of the stored quantum information at a desired time, e.g. for the basic task of synchronizing quantum repeater nodes across a quantum network~\cite{Wehner2018}. However, on-demand storage and retrieval is experimentally challenging at the single-photon level: laser pulses realizing the on-demand control generate optical noise, which limits the fidelity of the retrieved quantum information~\cite{Rakonjac2021}. On-demand operation also requires depleted auxiliary electronic levels, which limits the bandwidth of the memory and consequently the number of temporal modes \cite{Ortu2022b}. Hence, the record number of stored modes at the single-photon level is only 30 with on-demand operation~\cite{Ortu2022b}, which is in stark contrast to 1650 modes stored without on-demand read-out~\cite{Wei2024}.

In this paper, we realize an array of ten on-demand solid-state quantum memories, each individually controllable. We implement in each cell the full atomic-frequency comb (AFC) procotol~\cite{Afzelius2009}, which allows us to store single-photon-level light in independent spatio-temporal modes. We first store a total of $60$ modes, performed with an excited state storage time of $\SI{10}{\micro\second}$ and spin-wave storage time of $\SI{15.5}{\micro\second}$. We then increase the storage time in the excited state to $\SI{25}{\micro\second}$ to accommodate for a total of $250$ spatio-temporal modes, an order of magnitude more than the state of the art for solid-state spin-wave memories. We characterize the efficiency of each memory cell and the signal-to-noise ratio (SNR) of quantum storage with input pulses at the single-photon level. The average SNR is $31(9)$ for storage in $60$ modes and $10(2)$ for storage in $250$ modes. When summing the detections per storage trial of each spatio-temporal mode we find the cumulative totals to be higher for a higher number of stored modes: the probability of successfully detecting a photon after storing in $250$ modes is then higher than for $60$ modes, though the efficiency of each individual mode is lower. We furthermore assess the cross talk of the array at the single-photon level and find an average cross talk of $2.9\;\%$ across the array, therefore certifying them as individual and independent quantum memories.

\section{Results}
\subsection{Experimental implementation}

\begin{figure}[htpb]

    \subfloat{
        \includegraphics[width=1\columnwidth]{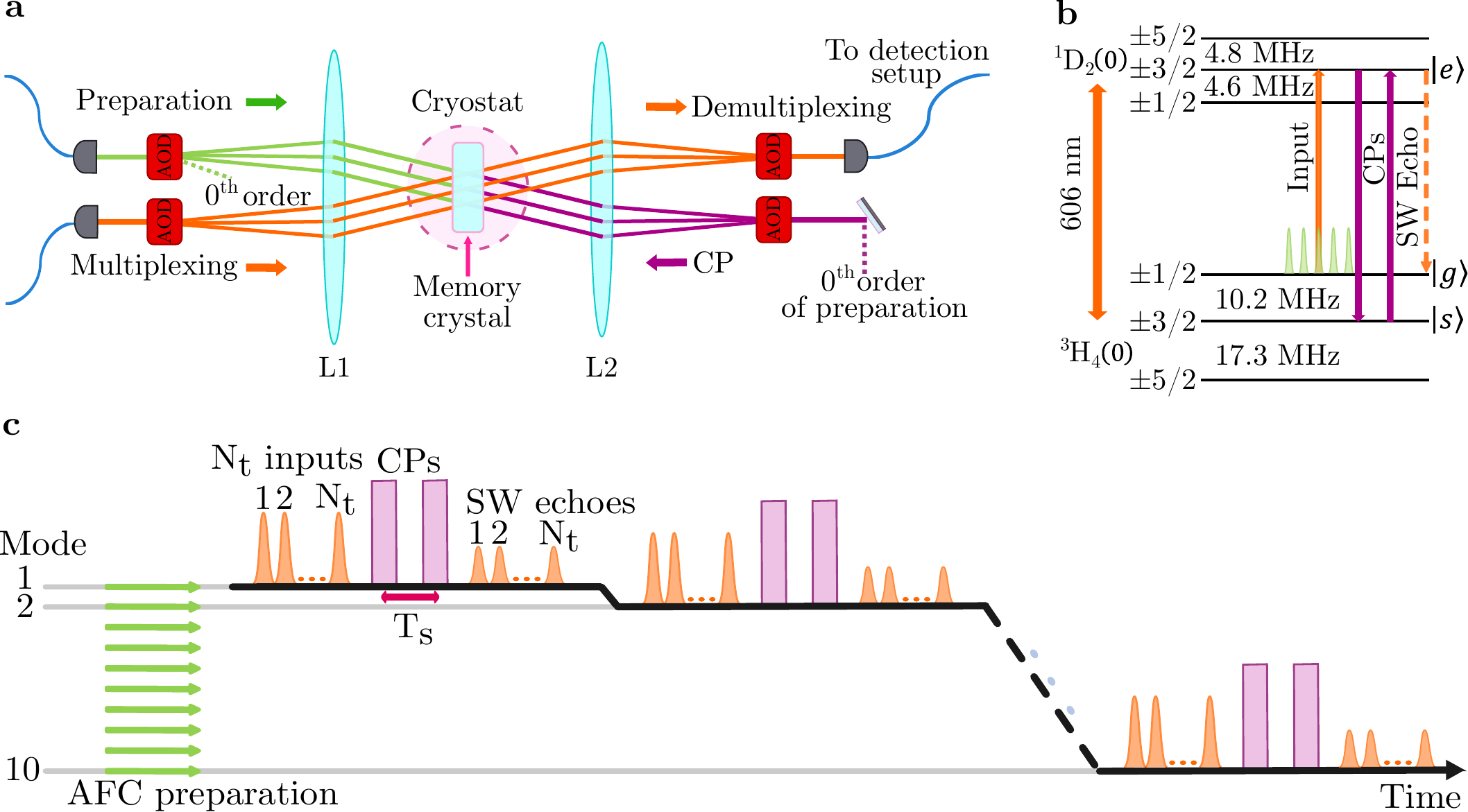}\label{fig1:setup_a}
    } 

    \subfloat{
        \includegraphics[width=0.0\columnwidth]{Figure1.pdf}
        \label{fig1:setup_b}
    }

    \subfloat{ 
        \includegraphics[width=0\textwidth]{Figure1.pdf}
        \label{fig1:sequence}
    }

    \caption{a) Schematic of the experimental setup. Acousto-optical deflectors (AODs) are used to multiplex and de-multiplex signal photons to and from individual quantum memory cells of a Pr$^{3+}:$Y$_2$SiO$_5$ crystal. The quantum memory is cooled to $\SI{3}{\kelvin}$ by a cryostat and is placed at distance $f=\SI{200}{\milli\meter}$ from lenses $L1$ and $L2$ with focal length $f$. Two additional AODs are used to select cells for their preparation and on-demand storage. Optical fibers guide the memory input, preparation, and control to the experimental setup and the retrieved signal to a filtering and detection setup. The control pulses are applied counter-propagating and co-linear to the preparation, to facilitate alignment and reduce the noise. b) Relevant electronic level structure of Pr$^{3+}:$Y$_2$SiO$_5$. An atomic frequency comb (AFC) is prepared using the sub-levels $\ket{g} = \pm1/2_g$ and $\ket{e} = \pm3/2_e$ of the $\SI{606}{\nano\meter}$ transition from $^{3}\mathrm{H}_4(0)$ to $^{1}\mathrm{D}_2(0)$ of \PrYSO. On-demand storage and retrieval is implemented with control pulses (CPs) on the transition from $\ket{e}$ to $\ket{s} = \pm 3/2_g$. c) Illustration of the experimental sequence. After simultaneous preparation of the memory array $N_\mathrm{t}$ input signals are sent into the first memory cell followed by a CP. After a storage time $T_s$, the stored signal is retrieved by a second CP. The sequence continues with $N_\mathrm{t}$ inputs into the next memory cell.}
\end{figure}

The optical setup of the quantum memory array is illustrated in Fig.~\ref{fig1:setup_a}. Two optical fibers guide light for the preparation and control pulses as well as the memory input to the experimental setup. Four acousto-optic deflectors (AODs) - one each for preparation of the memory array, multiplexing, on-demand control, and demultiplexing - are placed in pairs in the focal plane of two lenses $L1$ and $L2$ with focal length $f = \SI{200}{\milli\meter}$, with the multiplexing and demultiplexing AODs in a 4f configuration, with at the centre point a praseodymium-doped crystal in a cryostat. Here ten memory cells are identified by controlling the tones of the RF sent to the various AODs, with each cell having a $4\sigma$ diameter of $\approx \SI{80}{\micro\meter}$ with a separation of $\SI{200}{\micro\meter}$ between them. Light is stored in the array using the Atomic Frequency Comb (AFC) protocol~\cite{Afzelius2010a}. As visualised in Fig.~\ref{fig1:setup_b}, we prepare an AFC on the $\SI{606}{\nano\meter}$ transition from $\ket{g} = \pm1/2_g$ to $\ket{e} = \pm3/2$ between the states $^{3}\mathrm{H}_4(0)$ and $^{1}\mathrm{D}_2(0)$ of Pr$^{3+}:$Y$_2$SiO$_5$, simultaneously on all cells. In fact, the cells have a frequency separation of $\SI{1}{\mega\hertz}$, which is much smaller than the inhomogeneous broadening of \PrYSO$\;$and therefore does not influence the memory preparation.

Without on-demand storage, light absorbed by the comb is emitted after a fixed storage time $\tau$ determined by the separation between the teeth of the AFC. In the case of \PrYSO, the presence of three non-degenerate ground states allows us to leave one empty to implement on-demand storage and retrieval~\cite{Afzelius2009}. Frequency-selective and cell-specific control pulses (CP) can be applied using the relative AOD, transferring the atomic population from $\ket{e}$ to a ground-state spin-state $\ket{s} = \pm3/2_g$ where the AFC process stops. After a controllable spin-wave storage time $T_\mathrm{s}$ a second CP transfers the population from $\ket{s}$ back to $\ket{e}$, restarting the AFC echo process and leading to a coherent re-emission of the photon. The AFC protocol is temporally multimode, working in a first-in-first-out fashion for a maximum number of temporal modes given by the ratio between the mode duration and the AFC storage time~\cite{Ortu2022b}. Therefore, the realization of multiple quantum memory instances made possible by the AODs creates a spatially multiplexed array of temporally multiplexed memories.
These ten memory cells are demultiplexed using another AOD, which merges all the spatial modes back into one, which is finally coupled into a fiber connected to a detection setup with several filtering stages and a single-photon detector. For details on the detection setup see Sec.~\ref{Methods}. The demultiplexing AOD also deletes the frequency shift which was imparted to the light by the multiplexing AOD. Since the memory preparation sequence and CPs are never applied at the same time the respective AODs are never switched on simultaneously, which allows the 0th order light of the preparation AOD to be used for the CPs, further simplifying the setup.
The AODs are connected to an arbitrary-waveform generator (AWG), which provides us with precise control over the phase and amplitude of the light sent and retrieved from each cell. The ability to send several radio-frequency tones superimposed to the AODs allows us not only to prepare the AFC time-efficient in all cells simultaneously, but also to encode and store path qubits as well as storing time-bin qubits simultaneously in different cells. Experiments on the high-fidelity storage of path and time-bin qubits in the memory array are presented in Ref.~\cite{Teller2025}.

The storage sequence in the memory array is illustrated in Fig.~\ref{fig1:sequence} and begins with the preparation of the AFC simultaneously in all ten cells. Then, a variable number $N_\mathrm{t}$ of input signal pulses are sent into the first cell by the multiplexing AOD, followed by its cell-specific CP. Input pulses have a Gaussian profile with full-width at half maximum of $\SI{351(1)}{\nano\second}$. The control pulses are also Gaussian-shaped with a linear frequency chirp of $\SI{3.2}{\mega\hertz}$ and their total duration of $\SI{3.5}{\micro\second}$ is kept short not to limit the temporal multiplexing. After a variable storage time $T_s$ in the spin state, a second CP is applied on the same cell, and the stored light is emitted. The demultiplexing AOD then redirects it into the common spatial mode, which is then fibre coupled and sent to the detection setup. The sequence continues with $N_\mathrm{t}$ signal pulses into the next memory cell and repeats for the remaining memory cells. The memory preparation is performed once per cycle of the cryostat whereas the storage sequence is repeated 51 times. Light modulation for the memory preparation, input light and control pulses is realised using acousto-optic modulators controlled by the AWG.

\subsection{Characterization with strong light}

Before turning to the experiments at the single-photon level, we assess the efficiency of the memory array with classical light. We measure the efficiencies of the optical stages, including multiplexing, demultiplexing and output fiber coupling, and of the AFC and the two-way transfer efficiency, the combined efficiency to and from the spin state $\ket{s}$. We find that the efficiency of the multiplexing AOD ranges between $86(1)$ and $91.2(3)\;\%$. The de-multiplexing AOD shows a larger variation, with efficiencies between $65(3)$ and $92(8)\;\%$. The efficiency of the fiber coupling ranges from  $26(2)\;\%$ for the first cell to $58(3)\;\%$ for cell six. We attribute the discrepancy in the demultiplexing and the limited fiber efficiency to imperfect alignment of the $4f$ system and to optical aberrations introduced by the optical interfaces of the crystal. We emphasize that higher fiber coupling efficiencies uniform across the array may be achieved by carefully tilting the demultiplexing AOD~\cite{Pu2017}.

\begin{figure*}
    \centering
    \includegraphics[width=1\linewidth]{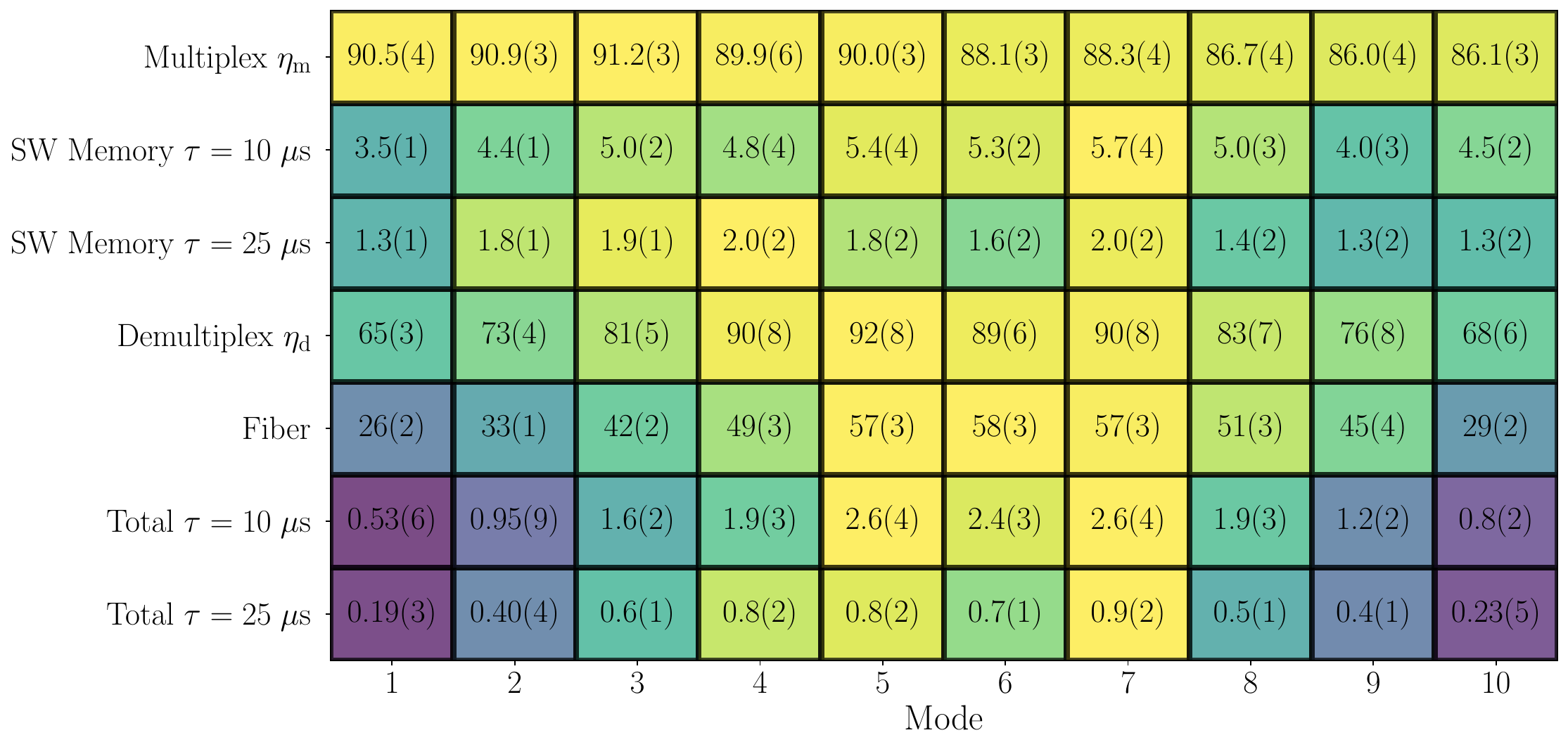}
    \caption{Efficiencies of the experimental setup obtained with classical light and listed in percent.}
    \label{fig:efficiencies}
\end{figure*}

We measure the efficiency for two AFC storage times $\tau$ of $10$ and $\SI{25}{\micro\second}$. For $\tau = \SI{10}{\micro\second}$ the AFC efficiency varies from $19.1(6)\;\%$ to $13.4(6)\;\%$. For $\tau = \SI{25}{\micro\second}$, the AFC efficiency spans from $7.9(4)$ to $4.0(3)\;\%$. The highest AFC efficiencies are comparable to experiments with a single spatial mode in \PrYSO~\cite{Rakonjac2021, Ortu2022b}. Lastly, we assess the combined efficiency of transferring to and from the spin-state $\ket{s}$ for a spin-wave storage time $T_s = \SI{15.5}{\micro\second}$ and find that it ranges from $20(1)$ to $36(4)\;\%$. Note that the two-way transfer is optimized for high signal-to-noise ratio at the single-photon level rather than for the highest efficiency. 
In fact, higher efficiencies for the spin-wave transfer have been observed with higher CP intensities, but resulted in a lower signal-to-noise ratio due to more fluorescence noise. Combining the transfer efficiency and AFC efficiencies results in spin-wave memory efficiencies between $3.5(1)$ and $5.7(4)\;\%$ for $\tau = \SI{10}{\micro\second}$ and between $1.3(1)$ and $2.0(2)\;\%$ for $\tau = \SI{25}{\micro\second}$. In summary, the total device efficiency of the multiplexed memory array ranges from $0.53(6)$ and $2.6(4)\%$ with an average of $1.6(2)\%$ for a storage time of $\tau = \SI{10}{\micro\second}$ and from $0.19(3)$ and $0.9(2)\%$ with an average of $0.6(1)\%$ for a storage time of $\tau = \SI{25}{\micro\second}$. The efficiencies of the multiplexing, the memories, demultiplexing, and fiber coupling are listed in Fig.~\ref{fig:efficiencies}. A list of all efficiencies including the AFC and two-way transfer is provided in Ref.~\cite{SuppMat_record}. 

\subsection{Storage of weak-coherent states}

We now turn to experiments at the single-photon level in which we store weak-coherent inputs with mean photon number $\bar{n} = 1.03(4)$. Following the sequence reported in Fig.~\ref{fig1:sequence}, we run it with two different configurations, with storage times of $\tau = \SI{10}{\micro\second}$ and $T_s = \SI{15.5}{\micro\second}$, and $\tau = \SI{25}{\micro\second}$ and $T_s = \SI{20}{\micro\second}$, storing, respectively, $60$ and $250$ spatio-temporal modes.  We then repeat the sequence without input pulses to measure the noise floor. Note that the mean photon number was calibrated after the multiplexing AOD and before the quantum memory array.

\begin{figure}[!htb]
    \centering
    \subfloat{
        \label{fig:Temp_a}
        \includegraphics[width=1\columnwidth]{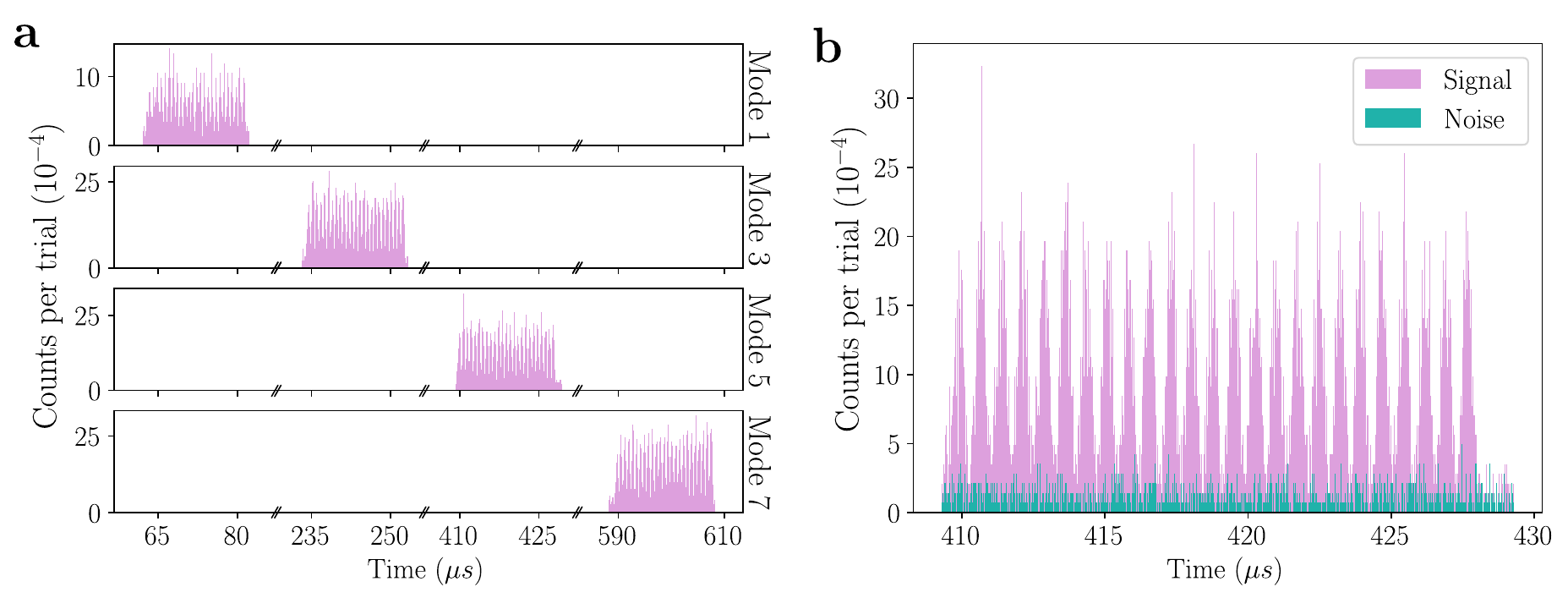}
    }
    \subfloat{
        \includegraphics[width=0.0\columnwidth]{Figure3.pdf}
        \label{fig:Temp_b}    
    }
     \caption{a) Detected counts per storage trial for $25$ temporal modes and four spatial modes. b) Detected signal and noise counts per temporal mode for mode five of a). }\label{Fig:Temp}
\end{figure}

\begin{figure}[!h]
    \centering
    \subfloat{
        \label{fig:2a}
        \includegraphics[width=1.00\columnwidth]{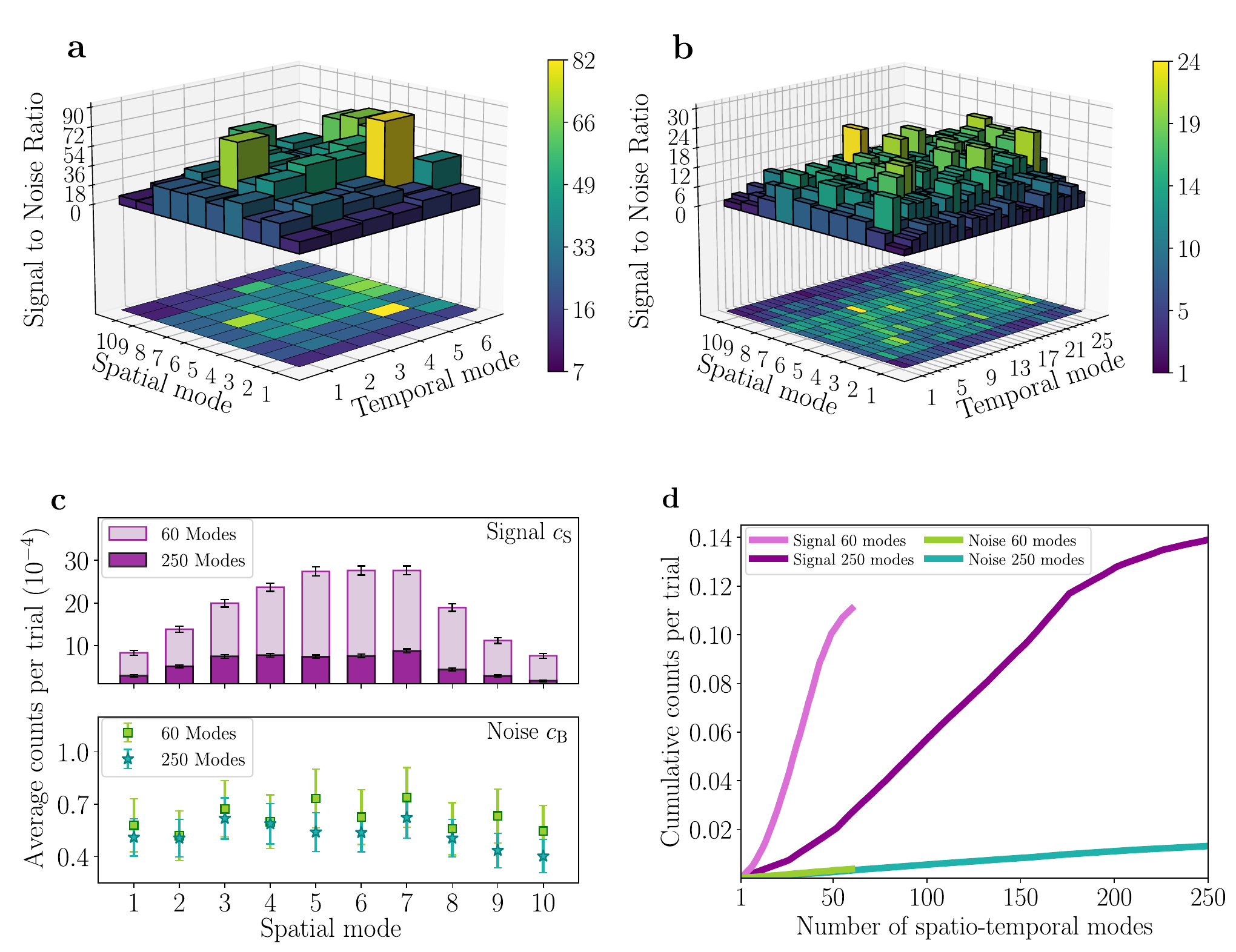}
    }
    \subfloat{
        \includegraphics[width=0.0\columnwidth]{Figure4.pdf}
        \label{fig:2b}    
    }

    \subfloat{
        \includegraphics[width=0.0\columnwidth]{Figure4.pdf}
        \label{fig:2c}
    }
    \subfloat{
        \includegraphics[width=0.0\columnwidth]{Figure4.pdf}
        \label{fig:2d}
    }

    \caption{Signal to noise ratio for $N_\mathrm{m}=60$ (a) and $N_\mathrm{m}=250$ (b) spatio-temporal modes. c) Signal counts $c_\mathrm{S}$ and background counts $c_\mathrm{B}$ per storage trial averaged over the $6$ and $25$ temporal modes of each spatial mode. d) Cumulative sum of detected signal and noise counts over $N$ spatio-temporal modes. }
\end{figure}

The retrieved signal of $25$ temporal modes averaged over $14227$ storage trials is plotted with the noise counts averaged over $14131$ trials for the spatial modes one, three, five, and seven in Fig.~\ref{fig:Temp_a}. A close-up on the traces of mode five is provided in Fig.~\ref{fig:Temp_b}. For both measurements with and without input signal we determine the detection events per storage trial in a $\SI{351}{\nano\second}$ window corresponding to the FWHM of the weak-coherent inputs and average over all trials. We determine the signal to noise ratio (SNR) for each spatio-temporal mode, which are plotted in Figs.~\ref{fig:2a} and ~\ref{fig:2b}. For 60 modes, the SNR ranges from $7(2)$ to $82(34)$ with an average of $31(9)$, and for $250$ modes from $1.2(5)$ to $24(6)$ with an average of $10(2)$. We find that the first temporal mode of each spatial mode has lower SNR due to additional fluorescence from the CPs. This fluorescence decays over time and thus temporal modes emitted later feature a lower noise floor. Also, we find that the spatial modes at the center of the array have a higher SNR, which is explained by the higher efficiencies of those memory cells, as visible in Fig.~\ref{fig:efficiencies}.

For further comparison, we average the detected events $c_S$ and the noise ones $c_B$ for each memory cell over all the temporal modes. The values are plotted in Fig.~\ref{fig:2c}: as expected from the variation in efficiency between the cells, the detected counts per trial vary for each spatial mode. Due to the difference in the AFC efficiencies, storing in 60 modes results on average in $3.3(3)$ times more detections per storage trial than storing in 250 modes. In contrast, with an average value of $1.2(1)$ times the detections per storage trial, the noise floor is constant across the memory cells, independent of the number of temporal modes and AFC storage time. To evaluate the potential gain in detected counts due to the spatially-multiplexed array, we sum the counts $c_{i}$ per storage trial in mode $i$ over $N$ spatio-temporal modes to determine the cumulative counts per storage trial
\begin{equation}
    c_N = \sum_{i=1}^{N}c_i \mathrm{.}
\end{equation}
We evaluate the cumulative detection counts $c_{\mathrm{S},N}$ and noise counts $c_{\mathrm{B},N}$ per trial in Fig.~\ref{fig:2d} for $N=1$ to $N=N_\mathrm{m}$, for the two storage configurations. As expected from the noise floor in Fig.~\ref{fig:2c}, the detected noise per trial $c_{\mathrm{B},N}$ increases linearly with the number of modes for both $N_\mathrm{m}=60$ and $250$. The cumulative counts per trial of signal photons $c_{\mathrm{S},N}$ increase with varying slopes due to the varying efficiencies across the array. Though the individual storage efficiency of each mode is higher for $N_\mathrm{m}=60$ than for $N_\mathrm{m}=250$, the cumulative counts per trial of $c_{\mathrm{S},250}=0.139(1)$ photons is higher than $c_{\mathrm{S},60}=0.111(1)$ for the lower storage time, highlighting the importance of multimodality for high-rate quantum communication~\cite{Chang2019, Li2021}.

\begin{figure}[!htp]
    \centering
    \subfloat{
    \includegraphics[width=0.7\columnwidth]{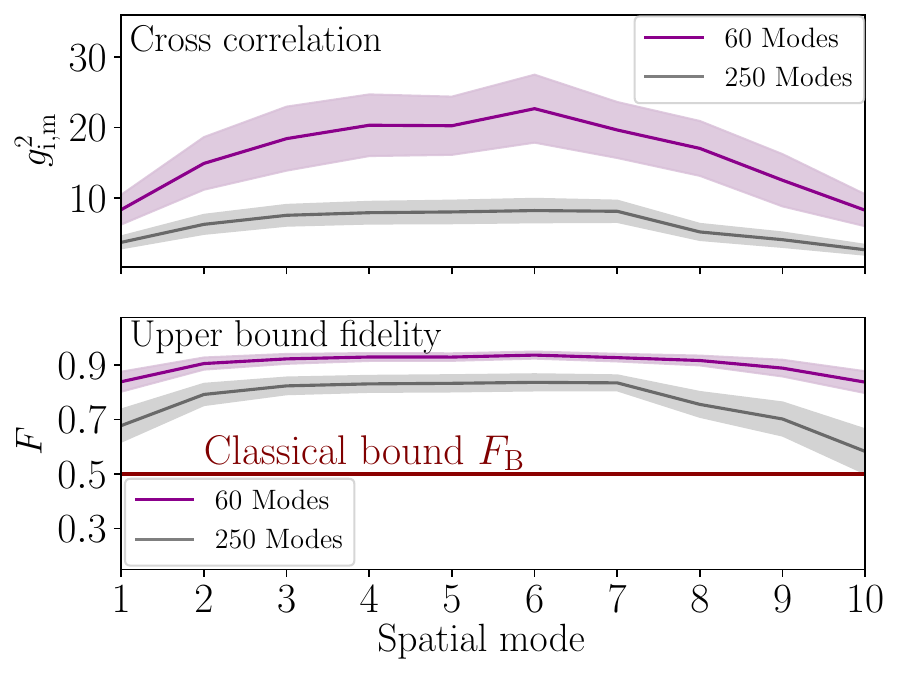}
    
    }
    \caption{Estimation of the performance of the memory array when storing non-classical states. a) Cross correlation $g^2_\mathrm{i,m}$ determined with Eq.~\ref{Eq:g2} from measured values of $SNR$ shown in Fig.~\ref{fig:2c} and literature values for $g^2_\mathrm{s,i}$~\cite{Rakonjac2021,Lago-Rivera2022}. b) Fidelity of an entangled state determined with Eq.~\ref{Eq:F}. }\label{fig:2e}
\end{figure}

Moving in this direction, we estimate the performance expected when deploying the memory array in a quantum network, starting from the measured SNR. In such a scenario, entangled pairs of non-degenerate photons will be used, with one photon at $\SI{606}{\nano\meter}$ and resonant with the memory cells, while the other at $\SI{1550}{\nano\meter}$ and compatible with telecom fibres~\cite{Lago-Rivera2021}. After absorption of the $\SI{606}{\nano\meter}$ photon, the memory and the telecom photon could not only exhibit non-classical correlations but could also be entangled, as the quantum memory preserves the correlations and entanglement if the SNR is sufficiently high~\cite{Rakonjac2021}. To consider a realistic photon-pair source, we assume that the heralding telecom photon has been generated with a corresponding $\SI{606}{\nano\meter}$ photon with a probability of  $\eta_\mathrm{H}=0.7$, which we further multiply by the multiplexing efficiency $\eta_\mathrm{m}$ per mode, as the single photons would travel through the multiplexing AOD.  Hence, we re-scale the average signal counts to $\Tilde{c}$ with
\begin{equation}
    \Tilde{c}_\mathrm{S} = c_\mathrm{S}\frac{\eta_\mathrm{m}\,\eta_\mathrm{H}}{\bar{n}} \mathrm{.}
\end{equation}
The re-scaled signal counts $\Tilde{c}_\mathrm{S}$ are about $\SI{60}{\percent}$ of the average counts per trial presented in Fig.~\ref{fig:2c}. With the adjusted signal to noise ratio 
\begin{equation}
    \Tilde{\textit{SNR}} = \frac{\Tilde{c}_\mathrm{S}-c_\mathrm{B}}{c_\mathrm{B}}
\end{equation}
we calculate the inferred second-order correlation function $ g^2_\mathrm{i,m}$ between the retrieved memory echo and the telecom photon~\cite{Albrecht2014}
\begin{equation}
    g^2_\mathrm{i,m} = g^2_\mathrm{s,i} \frac{\Tilde{\textit{SNR}} + 1}{g^2_\mathrm{s,i} + \Tilde{\textit{SNR}}} \label{Eq:g2}
\end{equation}
assuming a realistic value for the signal-idler cross-correlation between telecom and $\SI{606}{\nano\meter}$ photons of $g^2_\mathrm{s,i} = 100$~\cite{Lago-Rivera2021}. An upper bound on the fidelity of the entangled state with respect to the ideal state is then given by~\cite{deRiedmatten2005}
\begin{equation}
F = \frac{3}{4}\frac{g^2_\mathrm{i,m}-1}{g^2_\mathrm{i,m}+1}+\frac{1}{4} \mathrm{.}
\label{Eq:F}
\end{equation}
We plot the calculated correlation $g^2_\mathrm{i,m}$ with the extracted fidelity $F$ in Fig.~\ref{fig:2e}. Shaded areas indicate confidence intervals estimated from the uncertainties of the measured SNR. The simulated values of $g^2_\mathrm{i,m}$ range from $8(2)$ to $23(5)$ for $60$ modes and surpass the values for $250$ modes, which range from $2.6(7)$ and $8(2)$. With values as high as $94(1)\;\%$ for $N_\mathrm{m}=60$ and $84(3)\;\%$ for $N_\mathrm{m}=250$, the fidelities exceed the bound $F_\mathrm{B}=50\;\%$ for all the spatial modes. From the comparison with the classical bound, we conclude that the demonstrated SNR would be in principle sufficient to demonstrate quantum correlation and entanglement of each memory cell with a telecom photon~\cite{Rakonjac2021}. Our memory array could then be used efficiently and effectively in a quantum network, preserving quantum correlations between light and matter while increasing the communication rate through spatial and temporal multiplexing. 

\subsection{Cross talk at the single-photon level}

Finally, we asses the cross talk of the quantum memory array using weak-coherent inputs with mean photon number $\bar{n} = 0.95(3)$. We set a storage times $\tau = \SI{10}{\micro\second}$ and $T_s = \SI{8}{\micro\second}$ and perform spin-wave storage of a single input into spatial mode $i$ while selecting a mode $j$ with the demultiplexing AOD. We define cross talk $C$ between the input mode $i$ and the output mode $j$ as the ratio $C_{i,j} = c_{i,j}/c_{i,i}$ with $c_{i,j}$ denoting the detection events in mode $i$ per storage trial while collecting in mode $j$.
For comparison with previous experiments reported in Refs.~\cite{Lago-Rivera2021,Rakonjac2022,Lago-Rivera2023}, we shape the input signal to match the Lorentzian shape of single photons with a full-width at half maximum of $\SI{130(18)}{\nano\second}$ emitted by the cavity-enhanced spontaneous parametric down-conversion sources. 

\begin{figure*}[!htb]
    \includegraphics[width=0.95\columnwidth]{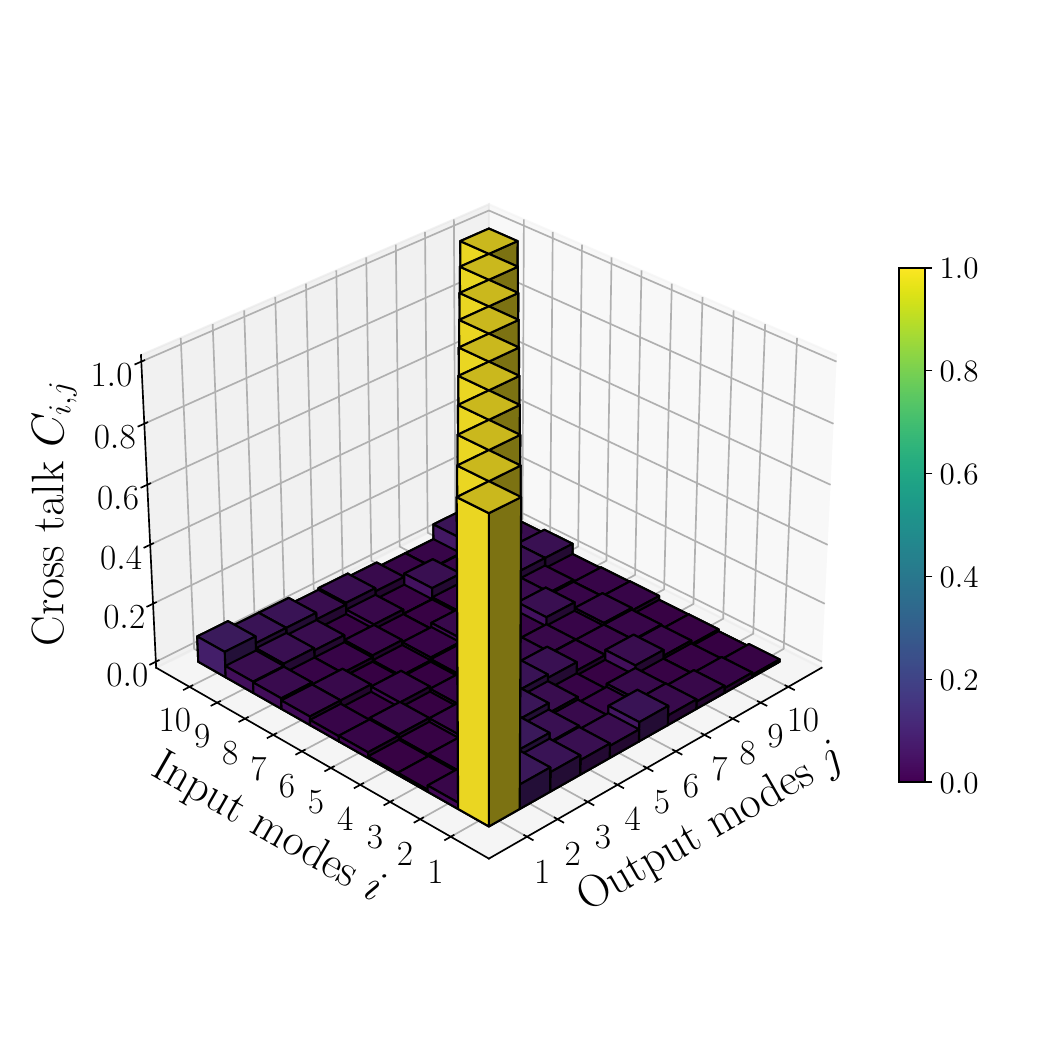}
    \caption{Cross talk $C_{i,j}$ determined from the ratio $C_{i,j} = c_{i,j}/c_{i,i}$ of detections per trial $c_{i,j}$ of input mode $i$ and output mode $j$. Errors are propagated from Poissonian statistics.} 
    \label{fig:cross talk}
\end{figure*}

We determine $C_{i,j}$ for all combinations of $i,j\in \{1,...,10\}$ considering a detection window of $\SI{351}{\nano\second}$, now containing $57\;\%$ of the photon, and the resulting cross talk matrix is plotted in Fig.~\ref{fig:cross talk}. We find that $C$ ranges from values as low as $0(1)\;\%$ up to $8.8(6)\;\%$ with an average cross talk of $2.9(1)\;\%$ across all combinations. We suspect that the origin of this cross talk is fluorescence noise due to the control pulses. While the collection fiber provides some spatial filtering, any imperfection in the alignment of the multiplexing-demultiplexing $4f$~system will let through some fluorescence noise, since it is emitted uniformly in space. To assess this hypothesis, we perform a storage sequence without input pulses and determine the noise counts per storage trial for the diagonal elements $n_{i,i}$. We calculate from this noise measurement the noise contribution $n_{i,i}/c_{i,i}$ to the cross talk, listed in Tab.~\ref{tab:CR}. We find these values to be comparable with the $C_{i,j}$ of the nearest neighbours in Fig.~\ref{fig:cross talk}. For modes further from the nearest neighbours, we find that both off-resonant AFC echos and two-pulse photon echoes of the CPs on the transition from $\ket{s}$ to $\ket{e}$ are shifted towards the resonance frequency of the memory by the demultiplexing AOD, and thus are transmitted through the filtering. This effect is in particular visible for combination of input mode $i=10$ with output modes $j=1,2,3$, bottom left in Fig.~\ref{fig:cross talk}. Note that the contribution from detector dark counts is on average $0.5(2)\;\%$ across the modes.

Future steps to reduce the cross talk will focus on improvements on the fiber coupling, increasing the SNR while decreasing the cross talk. Furthermore, with more efficient fiber coupling for outermost spatial modes we can increase the distance between the modes in the crystal, which will reduce fluorescence noise from the nearest neighbours coupled into the fiber. Moreover, with longer AFC storage times the resonant spin-wave echo can be positioned further away in time from the off-resonant AFC and two-pulse photon echos of the CPs, further reducing the noise.

An open study is the influence of the control fields manipulating a neighbouring cell $j$ on a cell $i$; a cross-talk study in this configurations is unfortunately not feasible with current storage times and switching times of the AODs (see Methods). With extended storage times, we will be able to conduct such a study by storing photons in a spatial mode $i$, subsequently applying several control pulses on a cell $j$, and then retrieving the stored light from the cell $i$. However, from a comparison of the $2\sigma$ radius of $\SI{50}{\micro\meter}$ of the control beams with the $\SI{200}{\micro\meter}$ separation between the memory cells, we expect the cross talk from the control fields to be minimal.

\begin{table}
    \centering
    \begin{tabular}{|c||c|c|c|c|c|c|c|c|c|c|}
    \hline
        Spatial mode & 1 & 2 & 3 & 4 & 5 & 6 & 7 & 8 &9 & 10\\
        \hline
        Cross talk $C_\mathrm{N} (\%)$  & 6(2) & 8(3) & 5(2) & 4(2) & 3(1) & 2(1) & 3(1) & 4(2) & 7(2) &10(2)\\
        \hline
    \end{tabular}
    \caption{Noise contribution $C_\mathrm{N}$ to the cross talk calculated from measurements of   $c_{i,i}$ without input.}
    \label{tab:CR}
\end{table}

\section{Discussion}
In conclusion, we realized a solid-state quantum memory array with ten individually controllable temporally multiplexed memory cells. We demonstrated on-demand storage at the single-photon level in 60 and 250 modes, surpassing the current state-of-the-art~\cite{Ortu2022b} for solid-state systems. The memory array showed a low cross-talk between the different cells, with an average of $2.9(1)\;\%$ across all spatial modes.
This combination of low cross-talk and high SNR allowed us in a separate study in Ref.~\cite{Teller2025} to store path and time-bin qubits with high fidelity. Given the current performances, we estimated that our system could be effectively deployed in a realistic scenario, preserving quantum correlations for all but two spatio-temporal modes.

In the near future, we will increase the overall detection probability by more than a factor three through improvements in the detection setup with superconducting nanowire detectors and more efficient fiber couplings. We will then connect this system to a cavity-enhanced photon-pair source and demonstrate non-classical correlations between the memory and the telecom photon. This approach to spatial multiplexing has the potential to significantly boost the entanglement generation rate in quantum repeaters, if the storage time can be increased. Combined with a photon pair source, it could then also serve as deterministic source of single and multiphotons states. The number of spatial modes can be further extended by using pairs of crossed AODs such that $100$ spatial modes are within reach~\cite{Lan2009, Pu2017}. Combining a 2D memory array with the storage of frequency-multiplexed light allowed by our solid-state quantum memory array~\cite{Seri2019} could lead to the combined storage of tens of thousands of modes in the spatial, spectral and temporal degree of freedom. Finally, this array may be combined with an impedance-matched cavity, which would allow for higher storage efficiencies exceeding $\SI{60}{\percent}$~\cite{Afzelius2010b}, similar to recent demonstrations of cavity-enhanced storage in cold-atomic quantum memory arrays~\cite{Wang2023}.

To increase the limited storage time, dynamical decoupling techniques~\cite{Viola1998} could be employed through either radio-frequency pulses acting globally on all memory cells simultaneously or through addressed optical pulses acting on individual memory cells. While the first approach has been shown to enhance the storage time of solid-state memories with a single spatial mode~\cite{Ortu2022a}, the timing of the decoupling pulses must be carefully chosen for the decoupling to work for all ten spatial modes at the same time. As an alternative, local optical pulses may be investigated as a path to realize optical dynamical decoupling for each spatial mode independently. Optimal control combined with pulse engineering techniques also may be applied to optimize the efficiency of these optical decoupling pulses~\cite{Borneman2010}.

\section{Methods}\label{Methods}
The AODs (DTSX-400-610 from AA Opto-electronic) are driven with amplified radio-frequency signals controlled by an AWG (HDAWG from Zurich Instruments). A pre-programmed sequence is uploaded to the AWG, which executes the storage sequence upon trigger from the cryostat cooling cycle. The switching times are measured to be $\SI{1.4}{\micro\second}$ for the AOD of the preparation beam, $\SI{2}{\micro\second}$ for the AOD of the control beam, $\SI{2.2}{\micro\second}$ for the multiplexing AOD and $\SI{2.3}{\micro\second}$ for the de-multiplexing AOD. The memory crystal consists a Y$_2$SiO$_5$ crystal matrix of size $\SI{5}{\milli\meter}\times\SI{2}{\milli\meter}\times\SI{5}{\milli\meter}$ with a dopant concentration of $0.05\;\%$. In the memory crystal, the $4\sigma$ diameters of the laser beams are approximately $\SI{200}{\micro\meter}$ for the preparation beam, $\SI{100}{\micro\meter}$ for the control fields, and $\SI{80}{\micro\meter}$ for the memory inputs. The light signal collected after the memory array is passing a filtering that consists of a filter crystal with a bandwidth of $\SI{4}{\mega\hertz}$, a narrow-band bandpass filter (Semrock Brightline FF01-600/14), and an etalon~\cite{Rakonjac2021}. Detections of the Laser Components COUNT-10FC single-photon detector during execution of the sequence are time-tagged and counted by the built-in counter of the AWG. This closed-loop approach allows in future for feed-forward operation and in-sequence control of stored quantum information. The total efficiency of the detection path is about $\SI{14}{\percent}$.

\section*{Acknowledgements}
This project received funding from: Gordon and Betty Moore Foundation (GBMF7446 to H. d. R); EU Horizon Europe Research and Innovation programme (EuroQCI in Spain) (Project no.101091638); Agència de Gestió d'Ajuts Universitaris i de Recerca; Centres de Recerca de Catalunya; FUNDACIÓ Privada MIR-PUIG; Fundación Cellex; Ministerio de Ciencia e Innovación with funding from European Union NextGeneration funds (MCIN/AEI/10.13039/501100011033, PLEC2021-007669 QNetworks, PRTR-C17.I1); Agencia Estatal de Investigación (PID2019-106850RB-100); Fundación Carmen y Severo Ochoa (BES-2017-082464, CEX2019-000910-S); European Union research and innovation program within the Flagship on Quantum Technologies through Horizon Europe project QIA-Phase 1 under grant agreement no. 101102140. M.T. acknowledges funding from the European Union's Horizon 2022 research and innovation programme under the Marie Sklodowska-Curie grant agreement No 101103143 "Two-dimensionally multiplexed on-demand quantum memories" (2DMultiMems). S.G. acknowledges funding from ``la Caixa'' Foundation (ID 100010434; fellowship code LCF/BQ/PR23/11980044). European Union’s Horizon 2020 Research and Innovation Programme under the Marie Skłodowska-Curie grant agreement number 956419 (NanoGlass).

This version of the article has been accepted for publication, after peer review but is
not the Version of Record and does not reflect post-acceptance improvements, or any corrections. The Version of Record is
available online at: \url{http://dx.doi.org/10.1038/s41534-025-01042-9}

\bibliography{qpsa.bib}
\newpage
\includepdf[pages=-]{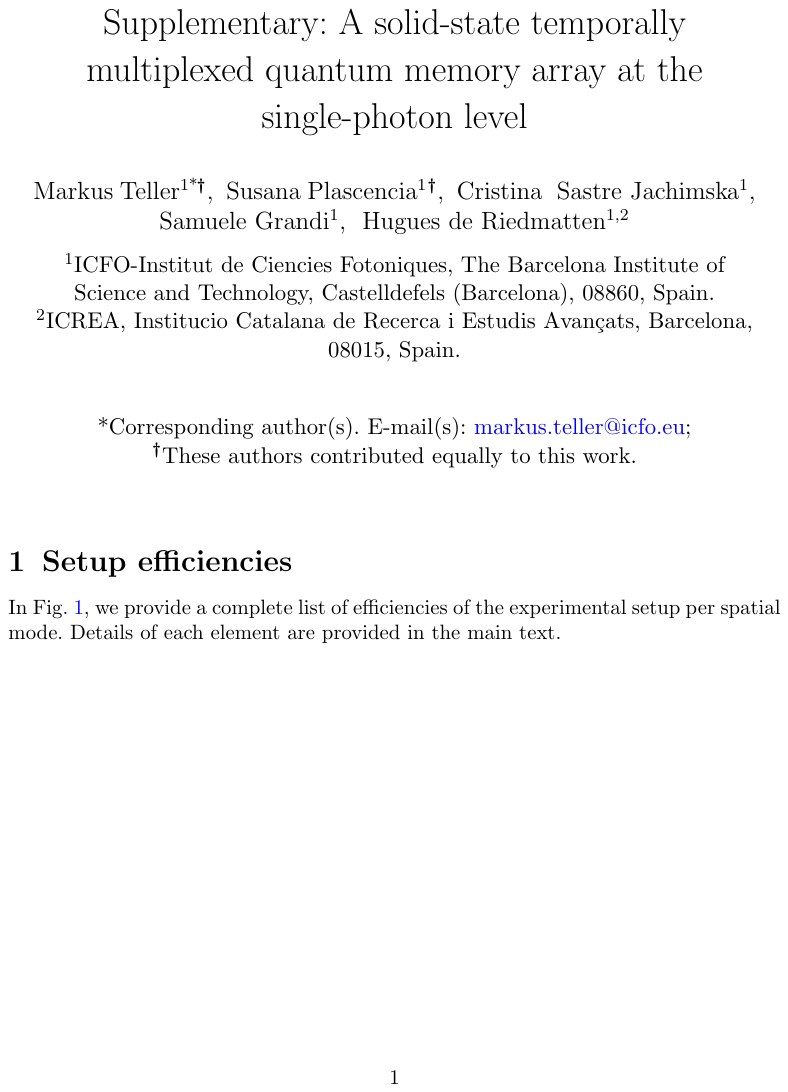}

\end{document}